\documentclass[prl,amsmath,superscriptaddress,twocolumn,floatfix]{revtex4}

\usepackage{graphicx}
\usepackage{amsmath}

\newcommand{\parti}[2]{\frac{\partial #1}{\partial #2}}

\newcommand{\diff}[2]{\frac{d #1}{d #2}}
\newcommand{\intall}{\int_{-\infty}^{\infty}}
\newcommand{\opex}{Opt. Express }

\newcommand{\bs}[1]{\boldsymbol{#1}}

\begin{document}
 \title{Coupled-Resonator Optical Near-Field
  Lithography}
\author{Mankei Tsang}
\email{mankei@optics.caltech.edu}
\affiliation{Department of Electrical Engineering,
  California Institute of Technology, Pasadena, California 91125, USA}
\author{Demetri Psaltis}
\affiliation{Department of Electrical Engineering,
  California Institute of Technology, Pasadena, California 91125, USA}
\affiliation{Institute of Imaging and Applied Optics, Ecole Polytechnique
 F\'ed\'erale de Lausanne, CH-1015 Lausanne, Switzerland}

\begin{abstract}
  The problem of pattern formation in resonantly-enhanced near-field
  lithography by the use of dielectric or plasmonic planar resonators
  is investigated. Sub-diffraction-limited bright or dark spots can be
  produced by taking advantage of the rotational invariance of planar
  resonators.  To increase the spatial bandwidth of the resonant
  enhancement, an array of coupled planar resonators can open up a
  band of Bloch resonances, analogous to coupled-resonator optical
  waveguides.
\end{abstract}

\maketitle

A major limitation of current near-field imaging systems is the
extreme proximity between the detector and the object. In lithography,
biological imaging, and optical data storage applications, any contact
with the lithographic mask, biological sample, or optical data storage
medium potentially causes damage and is undesirable in practice. The
detector is usually required to be close because the subwavelength
information of the object is carried by evanescent waves, which decay
exponentially. To increase the working distance, evanescent waves can
be amplified by passive optical resonators, such as dielectric slabs
\cite{tsang_ol,tsang_oe}, photonic crystals \cite{luo2}, plasmonic
slabs \cite{pendry,fang,melville}, and negative-refractive-index slabs
\cite{pendry}. The magnitude of the amplification is proportional to
$Q$, the quality factor of the resonance \cite{tsang_oe}, making the
use of low-loss dielectric structures most promising for near-field
imaging applications in the near future. Since the spatial bandwidth
of each resonance is inversely proportional to $Q$, however, a
resonator can only amplify certain sinusoidal patterns. An ideal
lossless negative-index slab would not suffer from this trade-off, but
such a material is impossible to fabricate by current technology.

\begin{figure}[htbp]
\centerline{\includegraphics[width=0.45\textwidth]{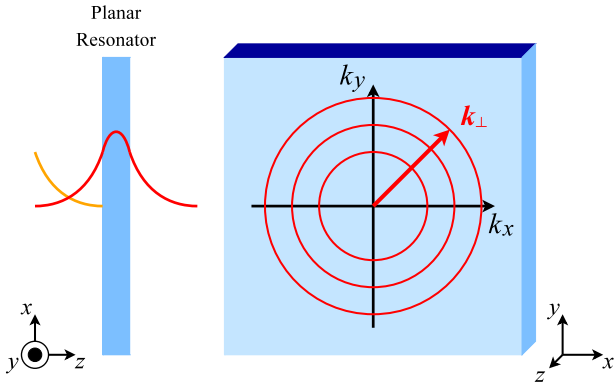}}
\caption{Left: Evanescent wave coupled into a resonance mode can be
  enhanced on the other side of the resonator. Right: For an infinite
  planar resonator, the resonance modes lie on circles in the
  transverse spatial frequency domain due to rotational invariance.}
\label{standing}
\end{figure}

Fortunately, complex patterns can still arise from the superposition
of sinusoidal waves in an ordinary resonator, thanks to
Fourier. Consider a planar resonator, such as a dielectric or
plasmonic slab, in the $x-y$ plane, as shown in
Fig.~\ref{standing}. An evanescent wave decaying in $z$ and impinging
upon the resonator can be amplified, if its transverse wave vector,
$\bs{k}_\perp \equiv k_x\hat{\bs{x}}+k_y\hat{\bs{y}}$, coincides with
that of a resonance mode. Because an infinite planar resonator is
rotationally invariant, the resonance condition depends on the
magnitude of $\bs{k}_\perp$ but not the direction of $\bs{k}_\perp$.
Hence, a planar resonator always possesses a continuum of azimuthally
degenerate modes. In practice, the finite size of the resonator means
that the system is not strictly rotationally invariant, but the
approximation is expected to be accurate for modes near the center of
a large resonator.

\begin{figure}[htbp]
\centerline{\includegraphics[width=0.45\textwidth]{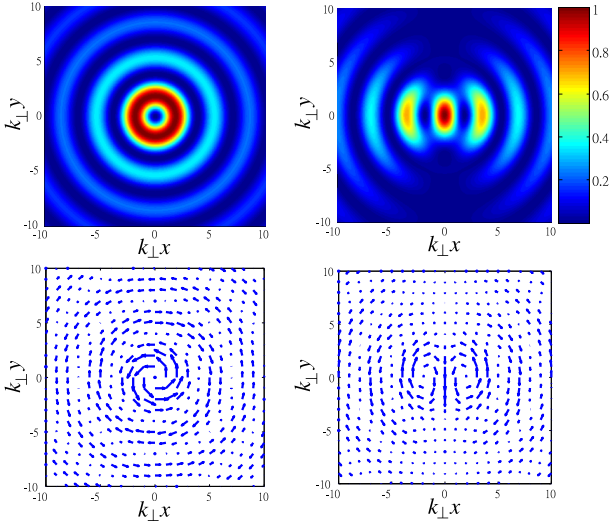}}
\caption{Free-space electric energy (top row) and electric field
  (bottom row) plots of the TE first-order Bessel mode (left column)
  and the dipole mode (right column).}
\label{TE_bessel}
\end{figure}

Exciting all the azimuthally degenerate modes produces evanescent
Bessel beams \cite{ruschin}, which can be especially useful for
lithography. The Bessel modes can be excited, for example, by placing
a solid immersion lens \cite{mansfield} near the resonator
\cite{tsang_oe}. For the transverse-electric (TE) resonance modes, the
free-space electric field in cylindrical coordinates $(\rho,\phi,z)$
is in general given by
\begin{align}
\bs{E}(\rho,\phi,z) &= \int_{-\pi}^\pi d\theta 
\left[\hat{\bs{\rho}}\sin(\phi-\theta)+
\hat{\bs{\phi}}\cos(\phi-\theta)\right]
\nonumber\\&\quad\times
f(\theta)\exp\left[ik_\perp \rho\cos(\phi-\theta)-\kappa z\right],
\end{align}
where $f(\theta)$ is an arbitrary complex function, $k_\perp$ is the
magnitude of $\bs{k}_\perp$ of a resonant mode, which must be
larger than the free-space wavenumber $k_0 \equiv \omega/c$, and
$\kappa \equiv (k_\perp^2-k_0)^{1/2}$ is the decay constant. Apart
from the exponential decay, the evanescent Bessel beam shapes remain
invariant along $z$, so one can put the detector or the photoresist
farther away from the resonator and still observe the same, albeit
attenuated, profile. Setting $f(\theta) = 1$, for instance, gives a
first-order Bessel beam,
\begin{align}
\bs{E}(\rho,\phi,z) &=
i\hat{\bs{\phi}}J_1(k_\perp \rho)\exp(-\kappa z),
\end{align}
which has a subwavelength dark spot in the middle. As the resonator is
also translationally invariant in the $x-y$ plane, we can
destructively interfere the first-order Bessel beam with one slightly
displaced in $x$ and obtain a ``dipole'' resonance mode,
\begin{align}
 \bs{E}(\rho,\phi,z)
&= i\parti{}{x}
\left[\hat{\bs{\phi}}J_1(k_\perp \rho)\right]\exp(-\kappa z),
\end{align}
which has two dark spots separated by a sub-diffraction-limited
distance of $3.68/k_\perp$.  The two examples of TE Bessel modes are
plotted in Fig.~\ref{TE_bessel}. Other resonant patterns are clearly
possible by specifying $f(\theta)$, or equivalently, superimposing
displaced Bessel modes.

\begin{figure}[htbp]
\centerline{\includegraphics[width=0.45\textwidth]{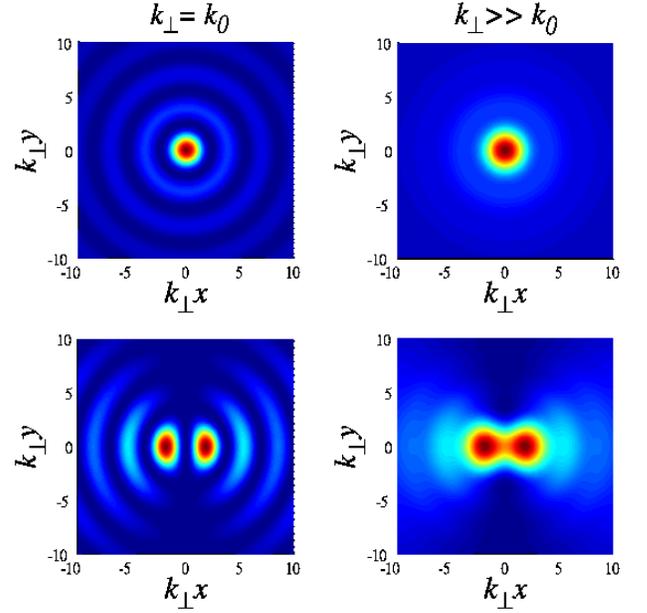}}
\caption{Free-space electric energy plots of the TM zeroth-order
Bessel mode (top row) and the dipole mode (bottom row) in the
diffraction limit (left column) and in the superresolution limit
(right column).}
\label{tmbessel}
\end{figure}

For the transverse-magnetic (TM) modes, on the other hand,
the free-space electric field is
\begin{align}
&\quad\bs{E}(\rho,\phi,z)
\nonumber\\ &= \int_{-\pi}^\pi d\theta 
\left[\hat{\bs{\rho}}\cos(\phi-\theta)-
\hat{\bs{\phi}}\sin(\phi-\theta)+
\hat{\bs{z}}\frac{ik_\perp}{\kappa}\right]
\nonumber\\&\quad\times
f(\theta)
\exp\left[ik_\perp \rho\cos(\phi-\theta)-\kappa z\right].
\end{align}
For $f(\theta) = 1$,
\begin{align}
\bs{E}(\rho,\phi,z) &= i\left[
\hat{\bs{\rho}} J_1(k_\perp \rho)
+\hat{\bs{z}}\frac{k_\perp}{\kappa}J_0(k_\perp \rho)\right]
\exp(-\kappa z).
\label{tmbesselfield}
\end{align}
The beam profile depends on $k_\perp$. In the diffraction limit,
$k_\perp = k_0$, the $\hat{\bs{\rho}}$ component vanishes, and the
beam profile contains a diffraction-limited bright spot. In the
superresolution limit, $k_\perp \gg k_0$, the total electric-field
energy is the incoherent superposition of the two polarizations, which
has a smoother beam shape but a slightly wider peak than the
zeroth-order Bessel function. A TM dipole mode can also be obtained by
differentiating Eq.~(\ref{tmbesselfield}) with respect to $x$.  The TM
Bessel modes are plotted in Fig.~\ref{tmbessel}.

Despite the presence of Bessel modes, near-field pattern formation
with just one value of $k_\perp$ is still fairly limited. In
particular, the side rings of Bessel modes are undesirable for
lithography. Here we show that an array of coupled planar resonators
can provide a band of resonance modes that help solve this problem.

\begin{figure}[htbp]
\centerline{\includegraphics[width=0.45\textwidth]{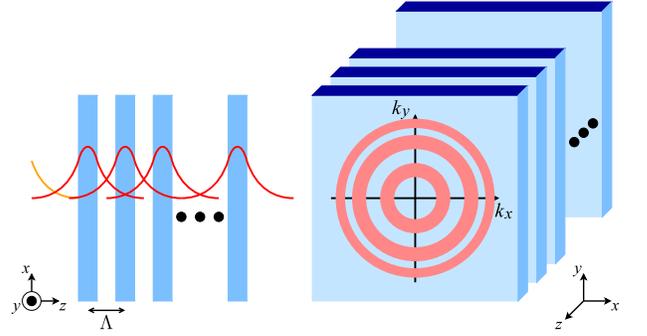}}
\caption{Coupled planar resonators possess Bloch resonances that
  enable evanescent-wave amplification for continuous bands of spatial
  frequencies.}
\label{crow_imaging}
\end{figure}

The physics of periodic layered media is well known
\cite{yeh_josa,yeh}. For simplicity, we study the problem in
terms of light propagation along $x$ in coupled waveguides, the
supermodes of which are identical to the resonance modes in the
evanescent coupling geometry in the limit of weak coupling and
lossless media, since the approximations used are identical in both
cases. Consider an array of $N$ identical planar resonators with
period $\Lambda$ along $z$, as depicted in Fig.~\ref{crow_imaging}. To
be specific, we shall consider the TE polarization only. Let the
permittivity of one resonator be $\epsilon'(z)$, and the electric
field be
\begin{align}
E_y(x,z) &=\sum_n A_n(x)\mathcal{E}_0(z-n\Lambda)\exp(ik_{\perp 0}x),
\end{align}
where $\mathcal{E}_0(z)$ is an unperturbed eigenmode electric field of
the resonator with $\epsilon'(z)$, normalized according to
$[k_{\perp 0}/(2\omega\mu_0)]\intall dz [\mathcal{E}_0(z)]^2 = 1$,
and $k_{\perp 0}$ is the spatial frequency along $x$ of the
unperturbed mode.  Invoking the weak coupling approximation, we obtain
\cite{yariv_qe}
\begin{align}
\diff{A_n}{x} &= i\beta A_n+i\gamma(A_{n-1}+A_{n+1}),
\label{coupledmode}
\end{align}
where $\beta \equiv (\omega/4) \intall dz [\epsilon(z)-\epsilon'(z)]
[\mathcal{E}_0(z)]^2$ and $\gamma \equiv (\omega/4) \intall dz
[\epsilon(z)-\epsilon'(z)] \mathcal{E}_0(z)\mathcal{E}_0(z-\Lambda)$.
The coupled-mode equations for the TM polarization are the same, with
slightly different definitions of $\beta$ and $\gamma$.  The $N$
coupled-mode equations (\ref{coupledmode}) give rise to $N$
supermodes. In the limit of infinite $N$, the supermodes become Bloch
modes,
\begin{align}
a_K &\equiv \sum_n A_n\exp(-inK\Lambda),
\\
a_K &\propto \exp[i\beta x + 2i\gamma\cos(K\Lambda)x],
\end{align}
which exist in a continuous band of spatial frequencies, with
a width controlled by the coupling constant $\gamma$,
\begin{align}
k_{\perp0} + \beta -2\gamma < k_\perp < k_{\perp0} + \beta +2\gamma.
\end{align}
The coupling constant can be manipulated by varying the distance
between each pair of resonators or changing the permittivity of the
gap material. In the evanescent coupling geometry, the Bloch modes
enable evanescent-wave amplification for a band of spatial frequencies
around the resonance of one planar resonator.

The physics of coupled planar resonators in the spatial domain is
analogous to the coupled-resonator optical waveguide (CROW) in the
time domain \cite{yariv_crow}, in which an array of coupled high-$Q$
resonators open up a band of resonant frequencies. Due to rotational
invariance, the Bloch theory is valid regardless of the direction of
$\bs{k}_\perp$, so the circles of resonances in the spatial frequency
domain for one resonator become thick rings for coupled resonators, as
depicted in Fig.~\ref{crow_imaging}. Furthermore, although our focus
here is monochromatic light, coupled planar resonators are also a CROW
in the time domain, and should therefore allow resonant transmission
of three-dimensional spatiotemporal information.

The magnitude of evanescent-wave amplification by each supermode can
be roughly estimated by comparing the energy of the supermode and the
power dissipation, using the method described in \cite{tsang_oe}. Let
$\Gamma$ be the evanescent-wave reflection coefficient of the
stack. When an incoming evanescent wave is resonantly coupled to the
structure, the stored energy in each supermode is approximately
proportional to $N\operatorname{Im}\{\Gamma\}^2$, while the power
dissipated at resonance is proportional to
$\operatorname{Im}\{\Gamma\}$. Using the definition of $Q$, the
maximum magnitude of $\Gamma$, after appropriate units are introduced,
is hence on the order of $Q/N$. The magnitude of the evanescent-wave
transmission coefficient is approximately the same as
$\operatorname{Im}\{\Gamma\}$ at resonance.

The possibility of anomalous diffraction due to Bloch dispersion or a
negative $\gamma$ may also be useful for imaging.  One way of
investigating anomalous diffraction is to take the limit of
infinitesimally thin layers and apply the effective medium theory
\cite{rytov,ramakrishna,salandrino,jacob,liu,smolyaninov,tsang_prb},
but the effective medium approximation is unable to account for
resonant enhancement. A Bloch theory in more complex geometries, such
as those studied in
Refs.~\cite{salandrino,jacob,liu,smolyaninov,tsang_prb}, will be of
fundamental interest. Nonlinear propagation in periodic media is
another interesting topic \cite{christodoulides}, although the problem
has not been studied in detailed in terms of the evanescent coupling
geometry. The new physics that arises from nonlinear optics may lead
to further improvement of resonantly enhanced near-field lithography.

Discussions with Zhiwen Liu, Andreas Vasdekis, James Adleman, and
Konstantinos Makris are gratefully acknowledged. This work is
supported by the National Science Foundation through the Center for
the Science and Engineering of Materials (DMR-0520565) and the DARPA
Center for Optofluidic Integration.


\begin{thebibliography}{}
\bibitem{tsang_ol}M.\ Tsang and D.\ Psaltis,
\ol \textbf{31}, 2741 (2006).

\bibitem{tsang_oe}M.\ Tsang and D.\ Psaltis,
\opex \textbf{15}, 11959 (2007).

\bibitem{luo2} C.\ Luo, S.\ G.\ Johnson, J.\ D.\ Joannopoulos,
and J.\ B.\ Pendry,
\prb \textbf{68}, 045115 (2003).

\bibitem{pendry}J.\ B.\ Pendry,
\prl \textbf{85}, 3966 (2000).

\bibitem{fang} N.\ Fang, H.\ Lee, C.\ Sun, and X.\ Zhang,
Science \textbf{308}, 534 (2005).

\bibitem{melville}D.\ O.\ S.\ Melville and R.\ J.\ Blaikie,
\opex \textbf{13}, 2127 (2005).

\bibitem{ruschin}S.\ Ruschin and A.\ Leizer,
\josaa \textbf{15}, 1139 (1998).

\bibitem{mansfield}S.\ M.\ Mansfield and G.\ S.\ Kino,
\apl \textbf{57}, 2615-2616 (1990).

\bibitem{yeh_josa}P.\ Yeh, A.\ Yariv, C.-S.\ Hong,
\josa \textbf{67}, 423 (1977).


\bibitem{yeh}P.\ Yeh,
\textit{Optical Waves in Layered Media}
(Wiley, New York, 2005).

\bibitem{yariv_qe}A.\ Yariv,
\textit{Quantum Electronics}
(Wiley, New York, 2001).

\bibitem{yariv_crow}A.\ Yariv, Y.\ Xu, R.\ K.\ Lee, and A.\ Scherer,
\ol \textbf{24}, 711 (1999).

\bibitem{rytov}S.\ M.\ Rytov,
Sov.\ Phys.\ JETP \textbf{2}, 466 (1956).

\bibitem{ramakrishna}S.\ A.\ Ramakrishna, J.\ B.\ Pendry,
M.\ C.\ K.\ Wiltshire, and W.\ J.\ Stewart,
\jmo \textbf{50}, 1419-1430 (2003).

\bibitem{salandrino}A.\ Salandrino and N.\ Engheta,
\prb \textbf{74}, 075103 (2006).

\bibitem{jacob}Z.\ Jacob, L.\ V.\ Alekseyev, and E.\ Narimanov,
\opex \textbf{14}, 8247 (2006).

\bibitem{liu}Z.\ Liu, H.\ Lee, Y.\ Xiong, C.\ Sun, and X.\ Zhang,
Science \textbf{315}, 1686 (2007).

\bibitem{smolyaninov}I.\ I.\ Smolyaninov, Y.-J.\ Hung, and
C.\ C.\ Davis,
Science \textbf{315}, 1699 (2007).

\bibitem{tsang_prb}M.\ Tsang and D.\ Psaltis,
\prb, in press.

\bibitem{christodoulides}D.\ N.\ Christodoulides, F.\ Lederer, and
Y.\ Silberberg,
\nat \textbf{424}, 817 (2003).
\end{thebibliography}
\end{document}